\documentclass[
]{ceurart}

\usepackage{xspace}
\usepackage{subcaption}

\begin{document}

\copyrightyear{2022}
\copyrightclause{Copyright for this paper by its authors.
  Use permitted under Creative Commons License Attribution 4.0
  International (CC BY 4.0).}

\conference{AIQxQIA 2026: International Workshop on AI for Quantum and Quantum for AI | co-located with AIxIA 2026, Perugia, Italy}

\newcommand{\bra}[1]{\langle #1 |}
\newcommand{\ket}[1]{| #1 \rangle}
\newcommand{\braket}[2]{\langle #1 | #2 \rangle}

\newcommand{\idest}{i.e.,\xspace}

\title{Reducing the Cost of Qudit State Preparation by Learning a Sparse Structure for SNAP Gates}

\author[1]{Marco Pennazio}[%
email=marco.pennazio@polimi.it,
orcid=0009-0000-2184-3929,
]

\author[1]{Leonardo Ignazio Pagliochini}[%
email=leonardoignazio.pagliochini@mail.polimi.it,
]

\author[1]{Riccardo Pellini}[%
orcid=0009-0005-0435-0264,
email=riccardo.pellini@polimi.it,
]

\author[1]{Maurizio {Ferrari Dacrema}}[%
orcid=0000-0001-7103-2788,
email=maurizio.ferrari@polimi.it,
]

\author[1]{Paolo {Cremonesi}}[%
orcid=0000-0002-1253-8081,
email=paolo.cremonesi@polimi.it,
]

\address[1]{Politecnico di Milano, Milano, Italy}

\begin{abstract}
Preparing quantum states in bosonic qudits with SNAP--Displacement sequences requires both optimizing the gate parameters and deciding which Fock levels should be controlled by each SNAP gate. A fully parameterized ansatz assigns an independent phase to every Fock level in every gate, leading to a number of phases that grows steeply. This constitutes a problem for short to medium term systems as controlling many phases requires complex control pulses which are subject to increased control errors. Previous work reduced this cost by restricting the trainable phases to predefined sparse patterns, such as a diagonal or a grid, which however may not be optimal for a specific target state. 
In this work, we address this limitation by introducing a method that learns which phases are needed together with their values, using a differentiable mask. A sparsity penalty encourages the removal of unnecessary phases, while temperature annealing gradually turns the masks into binary decisions. We compare the proposed method with fully parameterized and predefined sparse ansatzes on Fourier and Haar-random target states. The evaluation considers both the minimum number of phases required to reach different infidelity thresholds and the overall trade-off between fidelity and phases required, measured through Pareto frontiers and hypervolume. The results show that learning the location of the phases together with their values provides competitive or improved fidelity--resource trade-offs, although the advantage depends on the target state and qudit dimension.
\end{abstract}


\maketitle

\section{Introduction}
\label{sec:introduction}
Bosonic quantum computing encodes quantum information in superconducting radio-frequency cavities. These cavities can store different numbers of photons, with each photon-number level corresponding to a quantum state called a Fock state. While a qubit uses two levels, conventionally denoted by $\ket{0}$ and $\ket{1}$, a cavity can use many levels. By selecting the first $d$ Fock states $\ket{0} \ldots \ket{d-1}$, a single cavity can implement a $d$-level quantum system, or qudit \cite{SNAP2015PRL,PhysRevA.92.040303,chakram2022multimode}. These $d$ Fock states form a $d$-dimensional Hilbert space, which defines the computational space of the qudit. 
The potential advantages of encoding quantum information with qudits have been studied in many fields, including quantum simulation and quantum error correction. In particular, access to multiple levels can reduce the number of quantum devices required to represent the same information and, for some tasks, simplify the resulting circuits~\cite{10.3389/fphy.2020.589504,PRXQuantum.3.010335,PRXQuantum.4.020342,RevModPhys.97.021003}. While other technologies can also be used to create qudits, radio-frequency cavities offer both a large number of accessible Fock states and long coherence times~\cite{SNAP2015PRL,roy2024qudit,krayzman2024superconducting} with current implementations able to control the qudit up to $d=20$. 

Among the universal gate sets proposed for bosonic systems, a widely used one combines cavity displacement operations with Selective Number-dependent Arbitrary Phase (SNAP) gates~\cite{SNAP2015PRL,Heeres2017}. Displacements redistribute the population among the Fock levels, while SNAP gates apply independently controlled phases to the Fock levels. Alternating the two gates allows for universal control over the qudit and has been used for both state preparation and unitary synthesis~\cite{fosel2020efficient,PRXQuantum.3.030301,job2023efficient,roy2024qudit}. The gate parameters of a SNAP--Displacement sequence can be obtained through various methods: constructive compilation, numerical optimal control, or variational optimization~\cite{fosel2020efficient,job2023efficient,PRXQuantum.3.030301}. 
Preparing a target state with high fidelity remains a difficult problem. Constructive methods can decompose the preparation into a sequence of gates, but the number of gates can increase rapidly with the qudit dimension and the required fidelity. Variational methods can use shorter sequences, but a fully parameterized ansatz assigns one independent phase to each Fock level in every SNAP gate resulting in $Bd$ independent phase parameters for $B$ SNAP gates, which is a steep growth.

Besides increasing the time required to run the ansatz on the quantum device, a large number of phases creates an important drawback for practical applications. Controlling many Fock levels requires to engineer control pulses with several closely spaced frequency components, which can make SNAP gates longer and harder to calibrate~\cite{SNAP2015PRL,PRXQuantum.3.030301,roy2024qudit}, increasing the accumulation of control errors. Developing methods to identify the minimum number of phases required to prepare a state with the desired fidelity is therefore an active research area. 
Previous work addressed this problem by defining sparse SNAP--Displacement ansatzes in which only a subset of the phases can be used, according to heuristic patterns~\cite{dacrema2026sparse}. The \texttt{Diagonal (adaptive)}, \texttt{Diagonal (multiple)}, and \texttt{Grid} ansatzes were shown to reduce the number of active phases while maintaining high fidelity. However, the locations of the active phases are fixed before training and a pattern may retain phases that are not needed or exclude phases that would be useful for a specific target state.

In this work, we draw on differentiable masking methods from machine learning to overcome this restriction and make the selection of SNAP phases adaptive. The proposed \texttt{Learnable} method starts from the predefined \texttt{Diagonal (adaptive)} ansatz and associates each of the active phases with a trainable mask coefficient. These mask coefficients are optimized jointly with the phase parameters. A regularization term encourages the removal of unnecessary phases, while temperature annealing makes the learned mask progressively closer to a binary selection. The \texttt{Learnable} method therefore learns which phases within the starting ansatz should be retained for a given target state, as well as their optimal values.

We compare the proposed \texttt{Learnable} ansatz with the \texttt{Full}, \texttt{Diagonal (adaptive)}, \texttt{Diagonal (multiple)}, and \texttt{Grid} ansatzes using the same target states, qudit dimensions, and simulation setting. We first compare the minimum number of SNAP phases required to reach different infidelity thresholds. We then examine the Pareto frontiers to determine where each ansatz provides the best trade-off between infidelity and phase count, and use hypervolume to summarize these frontiers over both the complete explored range and the region satisfying $I\leq10^{-3}$. 
Overall, \texttt{Learnable} reduces the number of SNAP phases in several configurations and provides its clearest advantage for Fourier-5 at intermediate infidelities, while the results for Haar random states are more mixed.

\section{Background}
\label{sec:background}

The state preparation problem consists in finding a sequence of gates that transforms an initial state, typically $\ket{0}$, into a target state $\ket{\psi_{\mathrm{target}}}$. The quality of a prepared state $\ket{\psi_{\mathrm{prep}}}$ is measured through the fidelity $F=|\braket{\psi_{\mathrm{target}}}{\psi_{\mathrm{prep}}}|^2$, or equivalently through the infidelity $I=1-F$. A generic state of a $d$-dimensional qudit is a superposition of its $d$ Fock states, $\ket{\psi}=\sum_{f=0}^{d-1} c_f\ket{f}$, where $c_f\in\mathbb{C}$ and $\sum_{f=0}^{d-1}|c_f|^2=1$. In this paper we focus in particular on quantum states defined over a single qudit.

\subsection{The SNAP--Displacement Protocol}
\label{sec:snap_displacement}

The SNAP--Displacement protocol is a widely used approach for state preparation and unitary synthesis. It alternates displacement and SNAP gates to control the amplitudes and phases of the Fock states~\cite{SNAP2015PRL,Heeres2017}.

A displacement gate changes the population distribution $c$ across the Fock levels and is defined as:
\begin{equation}
    D(\alpha)
    =
    \exp\left(
        \alpha \hat{a}^{\dagger}
        -
        \alpha^* \hat{a}
    \right),
\end{equation}
where $\alpha\in\mathbb{C}$ is the displacement parameter. The operator $\hat{a}$ is called \emph{annihilation} operator and removes one photon from the cavity, acting as $\hat{a}\ket{f}=\sqrt{f}\ket{f-1}$ and $\hat{a}\ket{0}=0$, while the operator $\hat{a}^{\dagger}$ is called \emph{creation} operator and adds one photon $\hat{a}^{\dagger}\ket{f}=\sqrt{f+1}\ket{f+1}$. 

A SNAP gate applies an independent phase to each Fock state:
\begin{equation}
    S(\vec{\theta})
    =
    \sum_{f=0}^{\infty}
    e^{i\theta_{f}}
    \ket{f}\bra{f},
\end{equation}
where $\vec{\theta}=(\theta_0,\theta_1,\ldots)$ and $\theta_f\in\mathbb{R}$ is the phase applied to Fock level $f$.

Note that the Hilbert space of the cavity is infinite, while the hardware can directly control only a finite number of Fock levels, which correspond to the dimension $d$ of the qudit. However, displacement gates can also populate higher levels beyond what the hardware can control, and numerical simulations must take this effect into account to accurately represent the physical state of the cavity. This \emph{leakage} is typically undesirable because it moves part of the population outside the intended qudit, reducing the preparation fidelity. The simulations therefore include a finite number of additional levels above the first $d$, called \emph{bumper states}, which are not part of the qudit, but allow the simulation to account for population leaking outside it.

The SNAP--Displacement protocol consists of an alternating sequence of $B$ SNAP gates and $B+1$ displacement gates, and can be written as:
\begin{equation}
    U(\vec{\alpha},\Theta)
    =
    D(\alpha_B)
    S(\vec{\theta}_{B-1})
    D(\alpha_{B-1})
    \cdots  
    S(\vec{\theta}_0)
    D(\alpha_0),
\label{eq:sd_protocol}
\end{equation}
where $\vec{\alpha}=(\alpha_0,\ldots,\alpha_B)$ contains the displacement parameters. Each vector $\vec{\theta}_b$ contains the phase angles applied by the SNAP gate at block $b$, while $\Theta=[\theta_{b,f}]\in\mathbb{R}^{B\times d}$ collects the phases applied to the first $d$ Fock levels across the whole protocol. The phases of the bumper states are set to zero. Starting from $\ket{0}$, the state prepared by the protocol is
$\ket{\psi_{\mathrm{prep}}}=U(\vec{\alpha},\Theta)\ket{0}$.

In the fully parameterized ansatz, every entry of $\Theta$ is independently optimized. The protocol therefore contains $Bd$ SNAP phases and $B+1$ complex displacement parameters. Since this work studies how to reduce the resources required by SNAP gates, we use the number of active, \idest non-zero, SNAP phases as the main measure of the resources required.

The parameters of a SNAP--Displacement sequence can be determined through constructive compilation, numerical optimal control, or variational optimization~\cite{fosel2020efficient,job2023efficient,PRXQuantum.3.030301}. In this work, we follow the variational approach and optimize the displacement parameters and SNAP phases to maximize the fidelity with the target state.

\subsection{Predefined Sparse Phase Ansatzes}
\label{sec:predefined_sparse}
Previous work proposed reducing the number of SNAP phases by optimizing only those in specific locations~\cite{dacrema2026sparse}. The proposed patterns were motivated by the observation that, starting from $\ket{0}$, displacement gates tend to populate higher Fock levels progressively as the protocol advances. Therefore, phases applied to high Fock levels during the first blocks may have little or no effect because those levels have negligible population. 

The SNAP phases can be arranged as a matrix
$\Theta=[\theta_{b,f}]\in\mathbb{R}^{B\times d}$,
where the rows correspond to the protocol blocks and the columns to the Fock levels. A fixed binary \emph{structural mask}
$M^{\mathrm{struct}}\in\{0,1\}^{B\times d}$
controls which SNAP phases can be optimized, with the actual unitary used for state preparation being $U(\vec{\alpha},\widetilde{\Theta})$, with $\widetilde{\Theta}=M^{\mathrm{struct}}\odot\Theta$,
where $\odot$ denotes element-wise multiplication. Entries with
$M^{\mathrm{struct}}_{b,f}=1$ are learnable, while those with
$M^{\mathrm{struct}}_{b,f}=0$ are fixed to zero.

The following heuristic structures were considered in~\cite{dacrema2026sparse}:

\begin{itemize}
    \item \texttt{Full}: all phases are learnable, so that $M^{\mathrm{struct}}_{b,f}=1$ for every block and Fock level.

    \item \texttt{Diagonal (adaptive)}: the learnable phases lie within a band around the diagonal connecting the first and last Fock levels across the protocol. The diagonal follows
    $f=\frac{d-1}{B-1}b$, while a width hyperparameter determines the size of the band.

    \item \texttt{Diagonal (multiple)}: several parallel diagonal bands are used. Their number and width are set as hyperparameters.

    \item \texttt{Grid}: diagonal and anti-diagonal bands are combined, producing a grid of learnable phase locations. The number and width of the bands are set as hyperparameters.
\end{itemize}

These structures reduce the number of learnable phases, but the structural
mask is fixed once the heuristic and its hyperparameters have been
selected. The gate parameter optimization determines the values of the
learnable phases and may set some of them to zero, but it cannot change
which entries of $\Theta$ can be optimized. Our \texttt{Learnable} method which we introduce in the next section starts from the structural mask of the \texttt{Diagonal (adaptive)} ansatz and learns which phases within this mask should be retained.

\section{Learnable Phase Locations}
\label{sec:proposed_approach}

The heuristic ansatzes described in Section~\ref{sec:predefined_sparse} determine which SNAP phases are learnable before training. Our goal is to make this selection adaptive, so that the ansatz can retain only the phases that are useful for the target state rather than relying entirely on a predefined pattern. In our proposed \texttt{Learnable} method, we start from the mask defined by the \texttt{Diagonal (adaptive)} structure and introduce a second mask whose values are learned together with the gate parameters.

We refer to the mask defined by the \texttt{Diagonal (adaptive)} heuristic as the \emph{structural mask},
$M^{\mathrm{struct}}\in\{0,1\}^{B\times d}$.
This mask limits the phase locations considered by the method. A second mask, referred to as the \emph{learned mask}, $M^{\mathrm{learn}}\in[0,1]^{B\times d}$, determines which of these phases are retained. The two masks are combined through element-wise multiplication.

\subsection{Differentiable Learned Mask}
\label{sec:differentiable_mask}

A binary mask cannot be optimized directly using gradient-based methods, since changing one of its entries from zero to one is a discrete operation. We therefore use a continuous \emph{soft mask} during training and convert it into a binary mask only after the optimization is complete. For each entry for which $M^{\mathrm{struct}}_{b,f}=1$, we introduce a trainable logit $z_{b,f}$, which is an unconstrained real valued parameter that is mapped to the interval $[0,1]$ through a sigmoid function. The corresponding entry of the learned mask is defined as
\begin{equation}
    M^{\mathrm{learn}}_{b,f}
    =
    \sigma\left(
        \frac{z_{b,f}}{\tau}
    \right),
    \label{eq:soft_learned_mask}
\end{equation}

where $\sigma$ is the sigmoid function and $\tau>0$ is a temperature parameter. Positive values of $z_{b,f}$ favour retaining the phase, while negative values favour suppressing it. Similarly, the value of $M^{\mathrm{learn}}_{b,f}$ controls how much of the corresponding phase is retained, with values close to one leaving the phase almost unchanged, and values close to zero suppressing it. The structural and learned masks are combined into the effective mask
\begin{equation}
    M^{\mathrm{eff}}
    =
    M^{\mathrm{struct}}
    \odot
    M^{\mathrm{learn}},
    \label{eq:effective_mask}
\end{equation}
The effective protocol is therefore $U(\vec{\alpha},\widetilde{\Theta})$ where the SNAP phases used are
\begin{equation}
    \widetilde{\Theta}
    =
    M^{\mathrm{eff}}
    \odot
    \Theta.
    \label{eq:effective_learned_phases}
\end{equation}
If
$M^{\mathrm{struct}}_{b,f}=0$, the corresponding phase remains fixed to zero. If
$M^{\mathrm{struct}}_{b,f}=1$, its contribution is controlled by the learned mask.

Since the soft-mask values depend continuously on the logits, the logits can be optimized together with the gate parameters using gradient-based methods. The logits are initialized to $z_{b,f}=6$. With the initial temperature $\tau_{\mathrm{start}}=1$, this gives learned-mask values close to one, so the optimization starts from almost the complete structural mask.

During training, the temperature is reduced according to a geometric annealing schedule. At training step $t$, the temperature is
\begin{equation}
\label{eq:temperature_annealing}
    \tau_t
    =
    \tau_{\mathrm{start}}
    \left(
        \frac{\tau_{\mathrm{end}}}
             {\tau_{\mathrm{start}}}
    \right)^{q_t},
    \qquad
    q_t =
    \min\left(1,\frac{t}{N_{\mathrm{anneal}}}\right),
\end{equation}
where $N_{\mathrm{anneal}}$ is the number of steps over which annealing is performed. In our experiments, the temperature decreases from $\tau_{\mathrm{start}}=1$ to $\tau_{\mathrm{end}}=0.05$ over $10\,000$ training steps and remains fixed afterward.
At the beginning of training, the higher temperature gives a smoother dependence of the mask on the logits. As the temperature decreases, the sigmoid becomes steeper and the mask values move closer to zero or one. This avoids making sharp decisions before the phase and displacement parameters have had time to adapt. Annealing sharpens the mask values, while the joint objective introduced in Section~\ref{sec:joint_parameter_optimization} determines whether retaining or suppressing each phase is preferred.

At the end of training, the learned soft mask is converted into a binary hard mask using a threshold $\gamma$:
\begin{equation}
    M^{\mathrm{bin}}_{b,f}
    =
    \begin{cases}
        1, & M^{\mathrm{learn}}_{b,f} \geq \gamma,\\
        0, & M^{\mathrm{learn}}_{b,f} < \gamma.
    \end{cases}
\end{equation}

In our experiments, we use $\gamma=10^{-3}$. The final binary effective mask is
\begin{equation}
    M^{\mathrm{eff,bin}}
    =
    M^{\mathrm{struct}}
    \odot
    M^{\mathrm{bin}}
    \in\{0,1\}^{B\times d}.
    \label{eq:hard_effective_mask}
\end{equation}
The entries for which $M^{\mathrm{eff,bin}}_{b,f}=1$ identify the phases retained by the final mask. In order to mitigate the risk that the mask hardening process pushes the gate parameters in a suboptimal region, a short fine tuning phase of 200 steps is carried out using a learning rate reduced by a factor of ten. The final gate parameters are those used to compute the  fidelity between the prepared state and the target one.

The differentiable mask defines how phases can be retained or suppressed, but it does not by itself determine which choice is preferred. This is controlled by the joint optimization objective, which balances the fidelity of the prepared state with the number of the retained mask entries.

\subsection{Joint Parameter Optimization}
\label{sec:joint_parameter_optimization}

The displacement parameters, SNAP phases, and learned mask logits are optimized jointly. Following~\citet{dacrema2026sparse}, the gate parameters are obtained from unconstrained real-valued variables, which can be updated using gradient-based optimization. 
Each displacement gate is associated to a complex valued parameter $\alpha_b$ which is represented in polar coordinates using two real parameters $r_b,s_b\in\mathbb{R}$
\begin{equation}
    \alpha_b
    =
    \beta_b e^{i\varphi_b},
    \qquad
    \beta_b
    =
    \operatorname{ReLU}(r_b),
    \qquad
    \varphi_b
    =
    \pi\tanh(s_b).
    \label{eq:displacement_encoding}
\end{equation}
This mapping ensures that the displacement magnitude is non-negative and that its phase lies in $[-\pi,\pi]$. 
Similarly, each SNAP phase selected by the structural mask is obtained from an unconstrained parameter $t_{b,f}\in\mathbb{R}$ as
\begin{equation}
    \theta_{b,f}
    =
    \pi\tanh(t_{b,f}),
    \label{eq:snap_phase_encoding}
\end{equation}
which restricts the phase to $[-\pi,\pi]$. The logits $z_{b,f}$ introduced in Section~\ref{sec:differentiable_mask} are optimized together with these gate parameters.

The displacement parameters, SNAP phases, and mask logits are optimized by maximizing the fidelity of the prepared state while penalizing large gate parameters and the number of phases retained by the mask. The complete objective is
\begin{equation}
    \mathcal{J}
    =
    F\left(
        \vec{\alpha},
        \widetilde{\Theta}
    \right)
    -
    R_{\mathrm{gate}}
    -
    R_{\mathrm{mask}}.
    \label{eq:joint_objective}
\end{equation}

The first regularization term is applied to the gate parameters:
\begin{equation}
\begin{split}
    R_{\mathrm{gate}}
    ={}&
    \lambda_1
    \left(
        \lVert\vec{\alpha}\rVert_1
        +
        \lVert\widetilde{\Theta}\rVert_1
    \right) +
    \lambda_2
    \left(
        \lVert\vec{\alpha}\rVert_2^2
        +
        \lVert\widetilde{\Theta}\rVert_2^2
    \right).
\end{split}
\label{eq:gate_regularization}
\end{equation}
The $\ell_1$ term encourages some gate parameters to become zero, while the $\ell_2$ term limits large phase values and displacement amplitudes. We aim to set as many SNAP phase angles as possible to zero, as well as limiting large displacements because they can increase leakage to higher Fock levels and sensitivity to photon loss, which is a dominant noise mechanism in bosonic cavities. The weights $\lambda_1$ and $\lambda_2$ control the gate regularization. In the experiments for our \texttt{Learnable} method both are set to zero, so sparsity is controlled only through the learned mask. For the other baseline ansatzes, their values are instead searched with hyperparameter optimization as described in Section \ref{sec:baseline_optimization}. Note that the regularization is applied to the gate parameters, rather than to the unconstrained variables from which they are computed (see Eq. \ref{eq:displacement_encoding} and \ref{eq:snap_phase_encoding}), because only the gate parameters represent the physical operations that need to be performed.

The second regularization term controls the learned mask:
\begin{equation}
    R_{\mathrm{mask}}
    =
    \lambda_{\mathrm{mask}}
    \sum_{b,f}
    M^{\mathrm{eff}}_{b,f}
    +
    \lambda_{\mathrm{ent}}
    \sum_{b,f}
    H\left(
        M^{\mathrm{eff}}_{b,f}
    \right),
    \label{eq:mask_regularization}
\end{equation}
where $
    H(m)
    =
    -m\log m
    -(1-m)\log(1-m)
$ 
is the binary entropy, using the convention $0\log 0=0$, so that
$H(0)=H(1)=0$.

Since the effective mask is continuous during training, the sum
$\sum_{b,f}M^{\mathrm{eff}}_{b,f}$ provides a differentiable approximation of the number of phases retained by the mask. Minimizing this term favours mask values close to zero, while the fidelity term prevents phases that are needed for state preparation from being removed. The weight $\lambda_{\mathrm{mask}}$ therefore controls the trade-off between fidelity and the number of retained phases. 
The entropy term has a different role. Binary entropy is zero when a mask value is zero or one and is largest when it is equal to $0.5$. Penalizing this term therefore discourages intermediate values that do not favor retention or suppression. Together, the two mask terms encourage unnecessary phases to move towards zero and useful phases to remain close to one. The temperature schedule introduced in Section~\ref{sec:differentiable_mask} makes this separation progressively sharper.
Note that the mask regularization and the $\ell_1$ gate regularization act on different aspects of the ansatz. The mask regularization determines whether a phase remains learnable in the final ansatz, whereas the $\ell_1$ term acts on the numerical value of a phase and may set it to zero even when it is retained by the mask.

The unconstrained displacement variables, SNAP phase variables, and mask logits are updated jointly using Adam. The regularization weights $\lambda_1$, $\lambda_2$, $\lambda_{\mathrm{mask}}$, and $\lambda_{\mathrm{ent}}$ remain fixed during an individual training run and are selected through the hyperparameter optimization procedure described in Sections~\ref{sec:hyperparameter_optimization} and \ref{sec:baseline_optimization}.

After training, the hard effective mask from Equation~\ref{eq:hard_effective_mask} is applied. Entries equal to one identify the phases retained by the mask, while entries equal to zero fix the corresponding phases to zero. The gate parameters are then optimized for a further 200 steps, as described in Section~\ref{sec:training_protocol}. Greedy refinement and structural compression are subsequently applied as described in Section~\ref{sec:hyperparameter_optimization}.

\section{Methodology}
\label{sec:methodology}

\subsection{Experimental Setting}
\label{sec:experimental_setting}

We evaluate our proposed \texttt{Learnable} method on the same target states considered by~\cite{dacrema2026sparse}: \emph{Fourier-5} and \emph{Haar random} states defined for qudit dimensions $d\in\{8,16,32\}$. The SNAP--Displacement protocol starts from the vacuum state $\ket{0}$.
The \emph{Fourier-5} states are obtained by applying the $d$-dimensional Quantum Fourier Transform to the Fock state $\ket{5}$. The resulting state has equal amplitudes over the $d$ Fock levels and phases that vary regularly across them. The \emph{Haar random} target states are complex pure states sampled according to the Haar measure. 

Although the target states are defined over the first $d$ Fock levels, as we previously explained displacement operations can populate levels higher than $d$. The numerical simulations are performed in an extended Hilbert space containing
$n_{\mathrm{bumper}}=\max(64,2d)$ bumper states in addition to the $d$ levels of the qudit. The total dimension used in the simulations is
$\widetilde{d}=d+n_{\mathrm{bumper}}$. 
The target state is represented in the extended Hilbert space by assigning zero amplitude to all bumper states. Fidelity is evaluated over the complete space, so population leaking to the bumper states reduces the overlap with the target state.

We perform all simulations in double precision. The number of phases required by each method is evaluated at the target infidelity thresholds
$10^{-1}$, $10^{-2}$, $10^{-3}$, $10^{-4}$, and $10^{-5}$.

\subsection{Evaluation Protocol}
\label{sec:evaluation_protocol}

Our goal is to determine whether the \texttt{Learnable} ansatz can reach a given prepared state fidelity using fewer SNAP phases than the heuristic ansatzes, and whether this advantage remains consistent across a wide range of infidelity thresholds. The first stage is to identify the optimal hyperparameter configuration, as we describe in Sections~\ref{sec:hyperparameter_optimization} and \ref{sec:baseline_optimization}. Since a small number of random initializations can produce large outliers with very poor fidelity, we repeat the training over many random initializations and define the infidelity $I$ of each configuration as the mean infidelity over the $75\%$ of repetitions with the highest fidelity. Similarly, the number of phases associated to that hyperparameter configuration is the mean across the selected repetitions. These value are used throughout the analysis.
We first compare the ansatzes at a set of target infidelities. For each target state, qudit dimension, and threshold $\epsilon$, we consider all hyperparameter configurations satisfying $I\leq\epsilon$ and report the minimum number of SNAP phases among them. This analysis directly measures the number of phases required to reach a given preparation accuracy.

A reduction in the number of phases does not necessarily produce the same reduction in the physical duration of the ansatz, since the duration also depends on how the phases are distributed among the SNAP gates. We therefore report its estimated duration. Following~\citet{bornman2025benchmarking}, the duration of a SNAP gate in block $b$ is estimated as
\begin{equation}
    T_{S_b}
    \approx
    10\sqrt{p(\theta_b)}\,\mu\mathrm{s},
    \label{eq:execution_time}
\end{equation}
where $p(\theta_b)$ is the number of non-zero phase angles in that SNAP gate. The duration of a Displacement gate is much smaller and can be safely ignored. The ansatz duration is computed by adding the duration of all the SNAP gates in the ansatz. The duration associated to a hyperparameter configuration is the mean duration of the selected repetitions.

The analysis at fixed infidelity thresholds provides a direct comparison between ansatzes but only at those predefined thresholds. To provide a more complete picture, we also examine the Pareto frontiers. Each configuration is represented by a point $(p,I)$, where $p$ is the number of SNAP phases and $I$ is the aggregated infidelity. Both quantities should be minimized, so points closer to the lower-left region represent better trade-offs. A point belongs to the Pareto frontier when no other evaluated configuration has both a lower phase count and a lower infidelity, with at least one strict improvement. The frontier contains the configurations for which improving one objective requires worsening the other. It therefore shows the regions in which each ansatz provides the best balance between accuracy and phase count.

While the Pareto frontiers show where these differences occur and are useful
for visualization, comparing several frontiers across target states and qudit dimensions is not straightforward. We use the hypervolume indicator (HV) to summarize each frontier through a single value~\cite{hypervolume1,hypervolume2}. To compute the HV, a reference point is selected whose infidelity and number of phases are individually worse than those of all the evaluated configurations. Each point on the frontier defines a rectangular region extending towards this reference point, and the hypervolume is the area covered by the union of these regions. A larger hypervolume indicates a better trade-off, since it corresponds to lower infidelities, fewer phases, or both. For readability, the hypervolume values reported in the tables are multiplied by $100$.

We compute the hypervolume in two settings. The first includes the complete range of infidelities explored during the experiments and is directly comparable with~\cite{dacrema2026sparse}. The second includes only configurations satisfying $I\leq10^{-3}$ and focuses on the region where the Pareto frontiers show the main advantage of the \texttt{Learnable} ansatz. The normalization and reference point are computed independently for the two settings. Their absolute hypervolume values are therefore not compared across settings; comparisons are made only between ansatzes within the same setting and qudit dimension.

To quantify uncertainty, the reference point for each setting is computed using all repetitions and is then kept fixed during bootstrap resampling. For each hyperparameter configuration, the independent repetitions are sampled with replacement, and the mean infidelity is recomputed using the $75\%$ of repetitions with the highest fidelity. The Pareto frontier and its hypervolume are then recomputed. We repeat this procedure 1000 times and report the mean hypervolume and its $95\%$ confidence interval.

\subsection{Training Protocol}
\label{sec:training_protocol}
For each ansatz and hyperparameter configuration, the gate parameters are optimized through gradient ascent using Adam. In the \texttt{Learnable} ansatz, the mask logits are optimized jointly with the displacement parameters and SNAP phases according to the objective defined in Section~\ref{sec:joint_parameter_optimization}. The learning rate and regularization weights remain fixed within each training run. Their values are either fixed as part of the experimental setting or selected through the hyperparameter optimization procedures described in Sections~\ref{sec:hyperparameter_optimization} and~\ref{sec:baseline_optimization}.

Each training run is limited to $10^4$ optimization steps. Early stopping is applied when the best fidelity does not improve by at least $10^{-6}$ for $10^3$ consecutive steps. Reaching the target infidelity does not terminate the training run, since the threshold is used by the outer hyperparameter optimization to evaluate the final configuration. Since the optimization is sensitive to the initial gate parameters, each configuration is trained from multiple independent random initializations, also referred to as multistart training. Each baseline configuration is evaluated using 100 initializations, following~\cite{dacrema2026sparse}, while each \texttt{Learnable} configuration is evaluated using 10 initializations.

During each repetition, the parameter configuration achieving the highest fidelity is retained. For the \texttt{Learnable} ansatz, the temperature of the soft mask is updated at each optimization step using the schedule defined in Equation~\ref{eq:temperature_annealing}. After training, the soft mask is converted into the hard effective mask defined in Equation~\ref{eq:hard_effective_mask}. The retained gate parameters are then optimized for a further 200 steps using a learning rate reduced by a factor of ten, and the fidelity is recomputed using the resulting ansatz.

\subsection{Hyperparameter Optimization}
\label{sec:hyperparameter_optimization}
In a near term setting, the fidelity that can be achieved in practice is limited not only by the expressivity of the ansatz, but also by noise and other experimental imperfections. Increasing the number of phases to improve the fidelity beyond the range relevant to the application may therefore be counterproductive and result in a longer and more complex ansatz which is more subject to noise and control errors. We assume that this range is known and represent it through a target infidelity $\epsilon$. The aim of the hyperparameter optimization is then to identify a sparse ansatz whose final infidelity satisfies $I\leq\epsilon$. In our experiments $\epsilon \in \{10^{-5},10^{-4},10^{-3}\}$.

A separate hyperparameter optimization is performed for each target state, qudit dimension, and value of $\epsilon$. The hyperparameters of the \texttt{Learnable} ansatz are selected sequentially, and each configuration considered during the search is evaluated using the training protocol described in Section~\ref{sec:training_protocol}. The procedure assumes that increasing the width $w$ of the structural mask generally improves fidelity, while increasing $\lambda_{\mathrm{mask}}$ generally removes more phases at the cost of fidelity. These relations are not guaranteed because every configuration requires a separate non-convex optimization.
 The search works as follows. First, we fix the number of blocks to $B=d$ in all experiments, so it is not included in the search. The remaining hyperparameters are selected in two stages:
\begin{enumerate}
    \item select the width $w$ of the structural mask;
    \item select the mask regularization weight $\lambda_{\mathrm{mask}}$.
\end{enumerate}

During the first stage, the learned mask is disabled. Starting from $w=1$, the width is increased until a value satisfying $I\leq\epsilon$ is found, or the maximum width of $10$ is reached. The interval between the largest tested width that does not satisfy the threshold and the smallest width that satisfies it is then refined through two binary search steps. The smallest tested width satisfying the target infidelity is selected.

During the second stage, the selected width is kept fixed and the learned mask is enabled. We do not apply regularization directly to the gate parameters, while the entropy regularization weight is fixed to $\lambda_{\mathrm{ent}}=0.1$. The mask regularization weight is the only regularization coefficient selected by this search. The search first evaluates $\lambda_{\mathrm{mask}}=0$, for which the learned mask is disabled and only the structural mask is used. Positive values are then tested starting from $10^{-5}$ and increasing by a factor of $10$ until the target infidelity is no longer satisfied or the maximum value of $1$ is reached. The interval between the largest accepted value and the first rejected value is refined through four binary search steps. A final comparison is performed between the selected value and values $2\%$ above and below it. The largest tested value satisfying $I\leq\epsilon$ is selected.
 
The gate parameters obtained from the width search are used to initialize each configuration evaluated during the search for $\lambda_{\mathrm{mask}}$. Each tested value of $\lambda_{\mathrm{mask}}$ therefore starts from the same gate parameters rather than from the result obtained for the preceding value. 

\begin{figure}[htbp]
    \centering
    \includegraphics[width=0.9\textwidth]{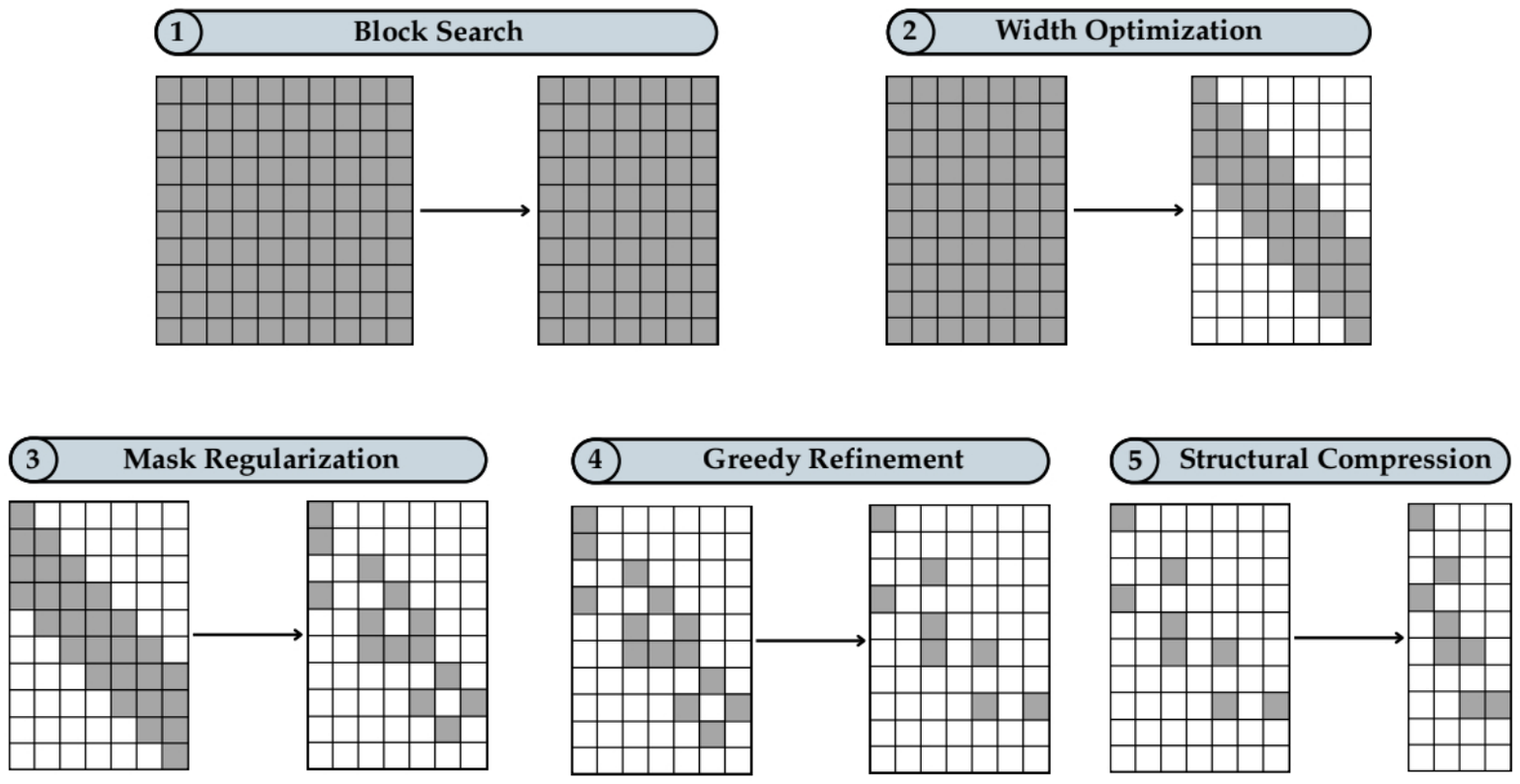}
    \caption{Overview of the optimization pipeline for the \texttt{Learnable} ansatz. In the experiments reported in this work, the number of blocks is fixed to $B=d$, so the block search shown in the first stage is omitted. The remaining stages include width selection, learned-mask optimization, greedy refinement, and structural compression.}
    \label{fig:training_process}
\end{figure}

After selecting $w$ and $\lambda_{\mathrm{mask}}$, the pipeline includes two final refinement steps. First, greedy refinement attempts to remove the remaining phases one at a time, accepting a removal only when the resulting ansatz still satisfies $I\leq\epsilon$. Accepted removals are retained during the subsequent tests. Second, structural compression removes any SNAP gate whose phase vector is entirely zero and combines its two adjacent displacement gates into a single displacement. Figure~\ref{fig:training_process} summarizes the complete procedure. The block search shown in the figure is omitted in our experiments, since we fix $B=d$.

Each hyperparameter optimization run is allocated a maximum wall time of four days. If the optimization does not complete within this limit, the search is terminated and the analysis uses all configurations completed before the interruption. This limitation mainly affects the experiments with $d=32$, for both the \emph{Fourier-5} and \emph{Haar random} states.

Unlike the baseline optimization described in Section~\ref{sec:baseline_optimization}, which uses different scalarization weights to explore the trade-off between fidelity and phase count, the sequential procedure searches directly for a sparse configuration satisfying a specified infidelity threshold. This keeps the search focused on the accuracy range relevant to the application.

\subsection{Baseline Hyperparameter Optimization}
\label{sec:baseline_optimization}
All baseline architectures (Full, Diagonal (adaptive), Diagonal (multiple), and Grid) are optimized following the optimization protocol used in \citet{dacrema2026sparse}, combining a scalarized loss and a Tree-structured Parzen Estimator (TPE) which is allowed to explore 50 cases.

The hyperparameter optimization of the baselines includes two sets of hyperparameters. 
The first controls the learning process, including the learning rate and the regularization weights $\lambda_1$ and $\lambda_2$. The second defines the structure of the ansatz and includes, for example, the number of blocks $B$, the diagonal width $w$, the number of diagonals $m_{\mathrm{diag}}$, and the number of anti-diagonals $m_{\mathrm{anti}}$, depending on the baseline architecture.

These hyperparameters, particularly those controlling the ansatz structure, affect both the fidelity of the prepared state and the required resources, such as the number of blocks and non-zero SNAP phases. To balance these competing objectives, \cite{dacrema2026sparse} formulated a multi-objective optimization problem that maximizes the fidelity while minimizing the number of non-zero SNAP phases. Following a standard scalarization approach~\cite{Bazgan2022}, the two objectives are combined into a single scalar objective used to evaluate each hyperparameter configuration $h$:

\begin{equation}
\label{eq:loss_hyp_ferrari}
L_h(\vec{\alpha}_h,\Theta_h,\beta)
=
F(\vec{\alpha}_h,\Theta_h)
-
\beta
\frac{p(\Theta_h)}{2d-2},
\end{equation}

where $F(\vec{\alpha}_h,\Theta_h)$ is the fidelity obtained after optimizing the gate parameters for a given hyperparameter configuration $h$, $p(\Theta_h)$ denotes the number of non-zero SNAP phase angles, and $\beta$ is a penalty coefficient controlling the trade-off between preparation fidelity and resource requirements. Its value is fixed before each hyperparameter optimization run and is not selected by TPE. The normalization factor $2d-2$ corresponds to the number of real degrees of freedom of a pure quantum state of dimension $d$, excluding normalization and global phase, allowing a consistent interpretation of $\beta$ across different qudit dimensions.

Note that the same paper \cite{dacrema2026sparse} also introduced a second scalarized loss penalizing the duration of the ansatz and observed that the duration is a much more difficult quantity to minimize. For this reason, in this study we focus on penalizing the number of phases.

Different values of $\beta$ produce different operating points, ranging from high-fidelity solutions requiring many non-zero SNAP phases to sparser solutions with lower resource requirements. We therefore perform a separate hyperparameter optimization for each value
\[
\beta\in
\{0,10^{-6},10^{-5},10^{-4},10^{-3},10^{-2},10^{-1}\}.
\]
For each value of $\beta$, the Tree-structured Parzen Estimator (TPE) explores at most 50 hyperparameter configurations. Using the same scalarization weights as~\cite{dacrema2026sparse} allows for a direct comparison between our \texttt{Learnable} method and the architectures they proposed.

\section{Results and Discussion}
\label{sec:results_and_discussion}

We evaluate the trade-off between preparation fidelity and the number of retained SNAP phases through three complementary analyses. First, for each target infidelity, we identify the ansatz requiring the minimum number of phases. Second, we compare the Pareto frontiers to examine the trade-off over the complete set of hyperparameter configurations explored. Third, we use hypervolume to summarize this trade-off both over the complete infidelity range and within the region satisfying $I\leq10^{-3}$, where the Pareto frontiers indicate \texttt{Learnable} is the most effective.

\subsection{Number of SNAP Phases per Target Infidelity}
\label{sec:phase_efficiency}

This first analysis compares the number of SNAP phases required by each ansatz to reach a given target infidelity. For each target state, qudit dimension, and threshold $\epsilon$, we consider all evaluated configurations satisfying $I\leq\epsilon$ and report the minimum number of SNAP phases. Lower values indicate that the same state preparation fidelity is reached using fewer phases, indicating the method exhibits a better trade-off.

As shown in Tables~\ref{tab:phase_fourier} and~\ref{tab:phase_rangauss}, \texttt{Learnable} requires fewer SNAP phases than all the baselines in 7 of the 15 Fourier-5 configurations. Its improvements are most consistent at $d=16$, where it performs best at three of the five infidelity thresholds. For the Haar random state, \texttt{Learnable} obtains 5 strict improvements and 2 ties. In this case, the strongest results are obtained at $d=8$, where it requires the fewest phases at four of the five thresholds, while the results become more mixed as $d$ increases.

For the Fourier-5 state, see Table~\ref{tab:phase_fourier}, \texttt{Learnable} is particularly effective for the larger qudit dimensions and the more demanding reachable thresholds. At $d=32$, it requires 81 phases for $I\leq 10^{-3}$ and 153 phases for $I\leq 10^{-4}$, compared with 152 and 348 phases, respectively, for the best baseline. These values correspond to reductions of approximately $47\%$ and $56\%$. A similar trend is observed at $d=16$, where \texttt{Learnable} reduces the phase count from 54 to 30 at $I\leq 10^{-3}$ and from 72 to 50 at $I\leq 10^{-4}$.

For the Haar random state, see Table~\ref{tab:phase_rangauss}, the improvements are more moderate and depend more strongly on the qudit dimension. At $d=8$, \texttt{Learnable} achieves the lowest phase count for thresholds between $10^{-2}$ and $10^{-4}$; in particular, it requires 16 phases at $I\leq 10^{-4}$, compared with 24 for the best baseline. At $d=16$ and the same threshold, the phase count decreases from 66 to 51. Conversely, \texttt{Learnable} is less competitive at $d=32$, where it does not achieve the minimum phase count at any of the reported thresholds. For example, at $I\leq 10^{-4}$, \texttt{Learnable} requires 153 phases, while the best baseline requires 142. \texttt{Learnable} also does not reach $I\leq 10^{-5}$ for $d=16$ and $d=32$ within the allotted computational budget of four days and the hyperparameter optimization is terminated before its completion.

The ansatz durations reported in the same tables show that a lower phase count can often be obtained without increasing the duration substantially. For the Fourier-5 state at $d=16$ and $I\leq 10^{-3}$, \texttt{Learnable} reduces the number of phases from 54 to 30, while its duration of $214\,\mu\mathrm{s}$ remains within $10\%$ of the shortest baseline duration of $195\,\mu\mathrm{s}$. For the Haar random state at $d=8$ and $I\leq 10^{-4}$, \texttt{Learnable} uses 16 phases and has a duration of $111\,\mu\mathrm{s}$, compared with 24 phases and $117\,\mu\mathrm{s}$ for the best baseline. The behavior is particularly favorable for the Fourier-5 state at $d=32$. At $I\leq 10^{-3}$, \texttt{Learnable} achieves both the lowest phase count and the shortest duration, requiring 81 phases and $495\,\mu\mathrm{s}$. At $I\leq 10^{-4}$, it requires 153 phases and $701\,\mu\mathrm{s}$, compared with 348 phases and $911\,\mu\mathrm{s}$ for the best baseline in terms of phase count. However, fewer phases do not always imply a shorter ansatz. For the Haar random state at $d=16$ and $I\leq 10^{-4}$, \texttt{Learnable} reduces the phase count from 66 to 51, but its duration increases from the best baseline value of $185\,\mu\mathrm{s}$ to $283\,\mu\mathrm{s}$. The least favorable case is observed for the Haar random state at $d=32$ and $I\leq 10^{-4}$, where \texttt{Learnable} minimizes neither the phase count nor the duration, requiring 153 phases and $701\,\mu\mathrm{s}$, compared with 142 phases and a minimum baseline duration of $283\,\mu\mathrm{s}$. This confirms that the ansatz duration depends not only on the total number of active phases, but also on their distribution across the SNAP gates and circuit blocks.

\begin{table*}[t]
    \centering
    \small
    \resizebox{0.95\textwidth}{!}{\begin{tabular}{cc|cc|cc|cc|cc|cc}
\toprule
 &  & \multicolumn{2}{c}{\texttt{Full}} & \multicolumn{2}{c}{\texttt{Diagonal (adaptive)}} & \multicolumn{2}{c}{\texttt{Diagonal (multiple)}} & \multicolumn{2}{c}{\texttt{Grid}} & \multicolumn{2}{c}{\texttt{Learnable}} \\
 \multicolumn{1}{c}{d} & \multicolumn{1}{c}{Infidelity $\leq$} \vline & Phases & T $[\mu s]$ & Phases & T $[\mu s]$ & Phases & T $[\mu s]$ & Phases & T $[\mu s]$ & Phases & T $[\mu s]$ \\
\midrule
\multirow[c]{5}{*}{8} & $10^{-1}$ & 8 & 28 & \textbf{2} & 20 & \textbf{4} & 40 & \underline{\textbf{1}} & 10 & \textbf{6} & 60 \\
 & $10^{-2}$ & 16 & 56 & \textbf{10} & 77 & \textbf{10} & 60 & \textbf{10} & 76 & \underline{\textbf{7}} & 70 \\
 & $10^{-3}$ & 24 & 85 & \textbf{16} & 88 & \textbf{15} & 92 & \textbf{16} & 87 & \underline{\textbf{8}} & 80 \\
 & $10^{-4}$ & 32 & 113 & \textbf{22} & 114 & \underline{\textbf{19}} & 106 & \textbf{20} & 113 & \textbf{21} & 127 \\
 & $10^{-5}$ & 40 & 141 & \textbf{28} & 128 & \underline{\textbf{22}} & 115 & \textbf{28} & 139 & \textbf{34} & 164 \\
\cline{1-12}
\multirow[c]{5}{*}{16} & $10^{-1}$ & 16 & 40 & \textbf{4} & 40 & \underline{\textbf{2}} & 20 & \textbf{3} & 24 & \textbf{14} & 135 \\
 & $10^{-2}$ & 48 & 120 & \textbf{16} & 143 & \textbf{21} & 149 & \textbf{26} & 155 & \underline{\textbf{14}} & 135 \\
 & $10^{-3}$ & 80 & 200 & \textbf{54} & 220 & \textbf{76} & 214 & \textbf{55} & 195 & \underline{\textbf{30}} & 214 \\
 & $10^{-4}$ & 80 & 200 & \textbf{72} & 253 & 87 & 228 & \textbf{77} & 255 & \underline{\textbf{50}} & 277 \\
 & $10^{-5}$ & 128 & 320 & \underline{\textbf{110}} & 346 & - & - & - & - & - & - \\
\cline{1-12}
\multirow[c]{5}{*}{32} & $10^{-1}$ & 64 & 113 & \underline{\textbf{10}} & 103 & \underline{\textbf{10}} & 77 & \textbf{12} & 68 & \textbf{32} & 323 \\
 & $10^{-2}$ & 96 & 170 & \underline{\textbf{34}} & 242 & \textbf{44} & 306 & \textbf{35} & 204 & \textbf{55} & 379 \\
 & $10^{-3}$ & 288 & 510 & \textbf{152} & 591 & \textbf{213} & 814 & \textbf{195} & 752 & \underline{\textbf{81}} & 495 \\
 & $10^{-4}$ & 544 & 963 & \textbf{348} & 911 & 663 & 1435 & 555 & 1304 & \underline{\textbf{153}} & 701 \\
 & $10^{-5}$ & - & - & - & - & - & - & - & - & - & - \\
% \cline{1-12}
% \multirow[c]{5}{*}{64} & $10^{-1}$ & 128 & 160 & \textbf{42} & 111 & \textbf{42} & 261 & \underline{\textbf{24}} & 68 & - & - \\
%  & $10^{-2}$ & 192 & 240 & \textbf{97} & 239 & \underline{\textbf{84}} & 218 & \textbf{106} & 184 & - & - \\
%  & $10^{-3}$ & 768 & 961 & \underline{\textbf{248}} & 754 & \textbf{394} & 1466 & \textbf{402} & 1456 & - & - \\
%  & $10^{-4}$ & 2880 & 3604 & \underline{\textbf{548}} & 1213 & \textbf{1067} & 2425 & \textbf{588} & 1895 & - & - \\
%  & $10^{-5}$ & - & - & \underline{2910} & 4267 & - & - & - & - & - & - \\
% \cline{1-12}
\bottomrule
\end{tabular}
}
    \caption{Comparison of the minimum number of non-zero phases required to reach different infidelity thresholds for the Fourier-5 target state.}
    \label{tab:phase_fourier}
\end{table*}

\begin{table*}[t]
    \centering
    \small
    \resizebox{0.95\textwidth}{!}{\begin{tabular}{cc|cc|cc|cc|cc|cc}
\toprule
 &  & \multicolumn{2}{c}{\texttt{Full}} & \multicolumn{2}{c}{\texttt{Diagonal (adaptive)}} & \multicolumn{2}{c}{\texttt{Diagonal (multiple)}} & \multicolumn{2}{c}{\texttt{Grid}} & \multicolumn{2}{c}{\texttt{Learnable}} \\
 \multicolumn{1}{c}{d} & \multicolumn{1}{c}{Infidelity $\leq$} \vline & Phases & T $[\mu s]$ & Phases & T $[\mu s]$ & Phases & T $[\mu s]$ & Phases & T $[\mu s]$ & Phases & T $[\mu s]$ \\
\midrule
\multirow[c]{5}{*}{8} & $10^{-1}$ & 16 & 56 & \underline{\textbf{6}} & 60 & \textbf{7} & 58 & \textbf{8} & 56 & \textbf{8} & 80 \\
 & $10^{-2}$ & 24 & 85 & \textbf{16} & 98 & \textbf{15} & 92 & \textbf{15} & 100 & \underline{\textbf{11}} & 87 \\
 & $10^{-3}$ & 32 & 113 & \textbf{19} & 104 & \textbf{19} & 106 & \textbf{22} & 120 & \underline{\textbf{15}} & 107 \\
 & $10^{-4}$ & 32 & 113 & \textbf{24} & 117 & \textbf{24} & 128 & \textbf{26} & 132 & \underline{\textbf{16}} & 111 \\
 & $10^{-5}$ & 40 & 141 & \underline{\textbf{34}} & 142 & \underline{\textbf{34}} & 154 & 45 & 164 & \underline{\textbf{34}} & 164 \\
\cline{1-12}
\multirow[c]{5}{*}{16} & $10^{-1}$ & 32 & 80 & \underline{\textbf{16}} & 131 & \textbf{21} & 134 & \textbf{20} & 106 & \underline{\textbf{16}} & 161 \\
 & $10^{-2}$ & 48 & 120 & \underline{\textbf{30}} & 155 & \textbf{37} & 148 & \textbf{35} & 169 & \textbf{32} & 223 \\
 & $10^{-3}$ & 64 & 160 & \textbf{44} & 161 & \textbf{57} & 182 & \textbf{54} & 241 & \underline{\textbf{42}} & 256 \\
 & $10^{-4}$ & 80 & 200 & \textbf{66} & 243 & \textbf{70} & 186 & \textbf{69} & 185 & \underline{\textbf{51}} & 283 \\
 & $10^{-5}$ & 112 & 280 & \underline{\textbf{98}} & 296 & - & - & \textbf{108} & 275 & - & - \\
\cline{1-12}
\multirow[c]{5}{*}{32} & $10^{-1}$ & 96 & 170 & \underline{\textbf{36}} & 274 & \textbf{49} & 240 & \textbf{38} & 212 & \textbf{72} & 452 \\
 & $10^{-2}$ & 96 & 170 & \underline{\textbf{64}} & 210 & 96 & 192 & \textbf{85} & 275 & \textbf{76} & 473 \\
 & $10^{-3}$ & 128 & 226 & \underline{\textbf{96}} & 238 & 129 & 251 & \textbf{127} & 225 & \textbf{121} & 614 \\
 & $10^{-4}$ & 160 & 283 & \underline{\textbf{142}} & 393 & 197 & 370 & 161 & 308 & \textbf{153} & 701 \\
 & $10^{-5}$ & 448 & 793 & 673 & 1428 & - & - & \underline{\textbf{320}} & 593 & - & - \\
% \cline{1-12}
% \multirow[c]{5}{*}{64} & $10^{-1}$ & 192 & 240 & \underline{\textbf{92}} & 429 & \textbf{112} & 473 & \textbf{106} & 440 & - & - \\
%  & $10^{-2}$ & 320 & 400 & \underline{\textbf{186}} & 358 & \textbf{207} & 284 & \textbf{219} & 324 & - & - \\
%  & $10^{-3}$ & 320 & 400 & \underline{\textbf{266}} & 538 & \textbf{292} & 412 & 351 & 494 & - & - \\
%  & $10^{-4}$ & 896 & 1121 & 1686 & 2586 & \underline{\textbf{799}} & 1175 & 954 & 1236 & - & - \\
%  & $10^{-5}$ & - & - & \underline{3570} & 4738 & - & - & - & - & - & - \\
% \cline{1-12}
\bottomrule
\end{tabular}
}
    \caption{Comparison of the minimum number of non-zero phases required to reach different infidelity thresholds for the Haar random target state.}
    \label{tab:phase_rangauss}
\end{table*}

\subsection{Pareto Frontiers}
\label{sec:pareto_frontiers}

The previous analysis considers a set of fixed infidelity thresholds. To examine the complete trade-off between preparation accuracy and the number of SNAP phases, we also compare the Pareto frontiers obtained by the different ansatzes. A configuration belongs to the Pareto frontier when no other evaluated configuration achieves both a lower infidelity and a lower number of phases.

\begin{figure}[t]
    \centering

    \begin{subfigure}[t]{0.48\linewidth}
        \centering
        \includegraphics[width=\linewidth]
        {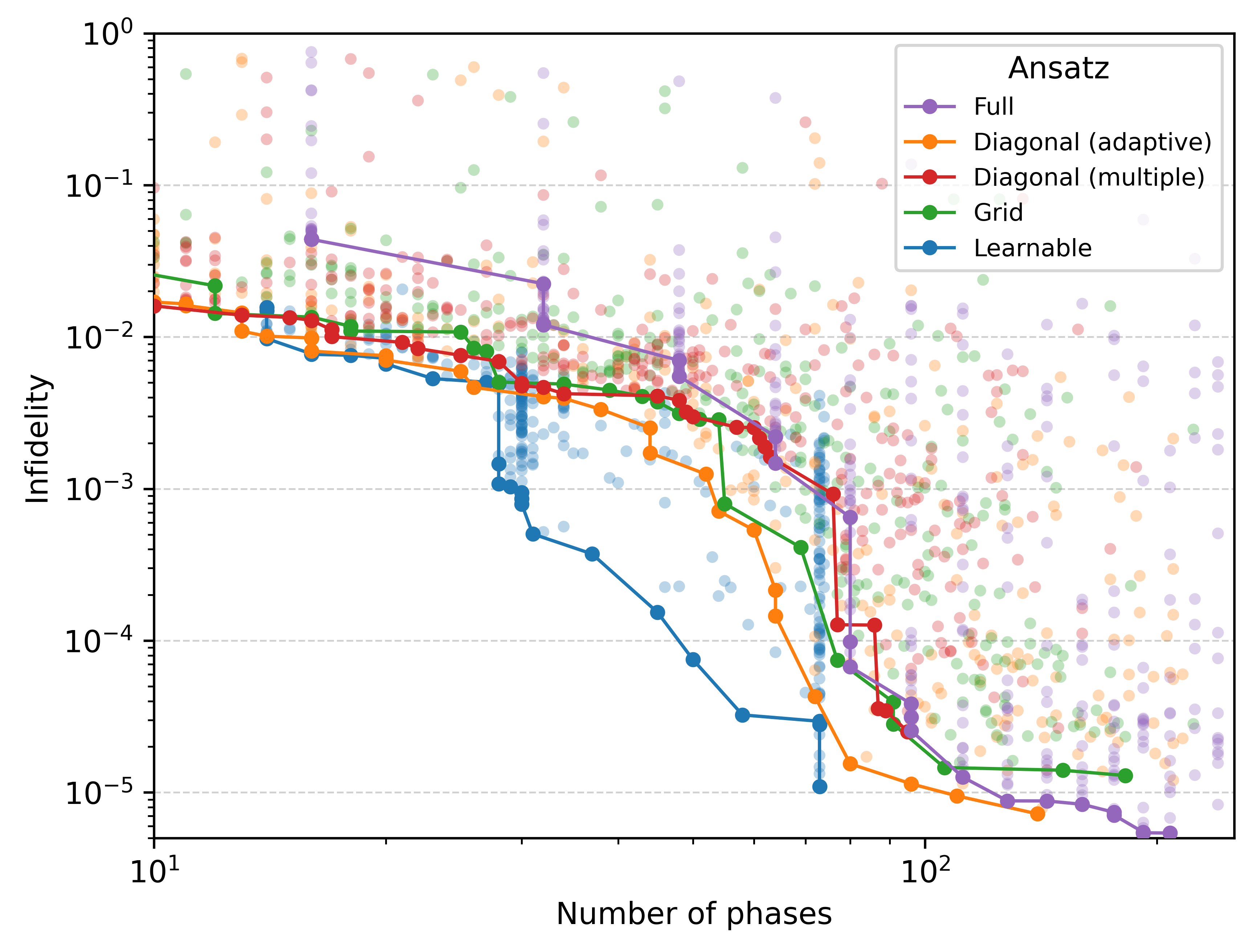}
        \caption{$d=16$.}
        \label{fig:pareto_fourier_d16}
    \end{subfigure}
    \hfill
    \begin{subfigure}[t]{0.48\linewidth}
        \centering
        \includegraphics[width=\linewidth]
        {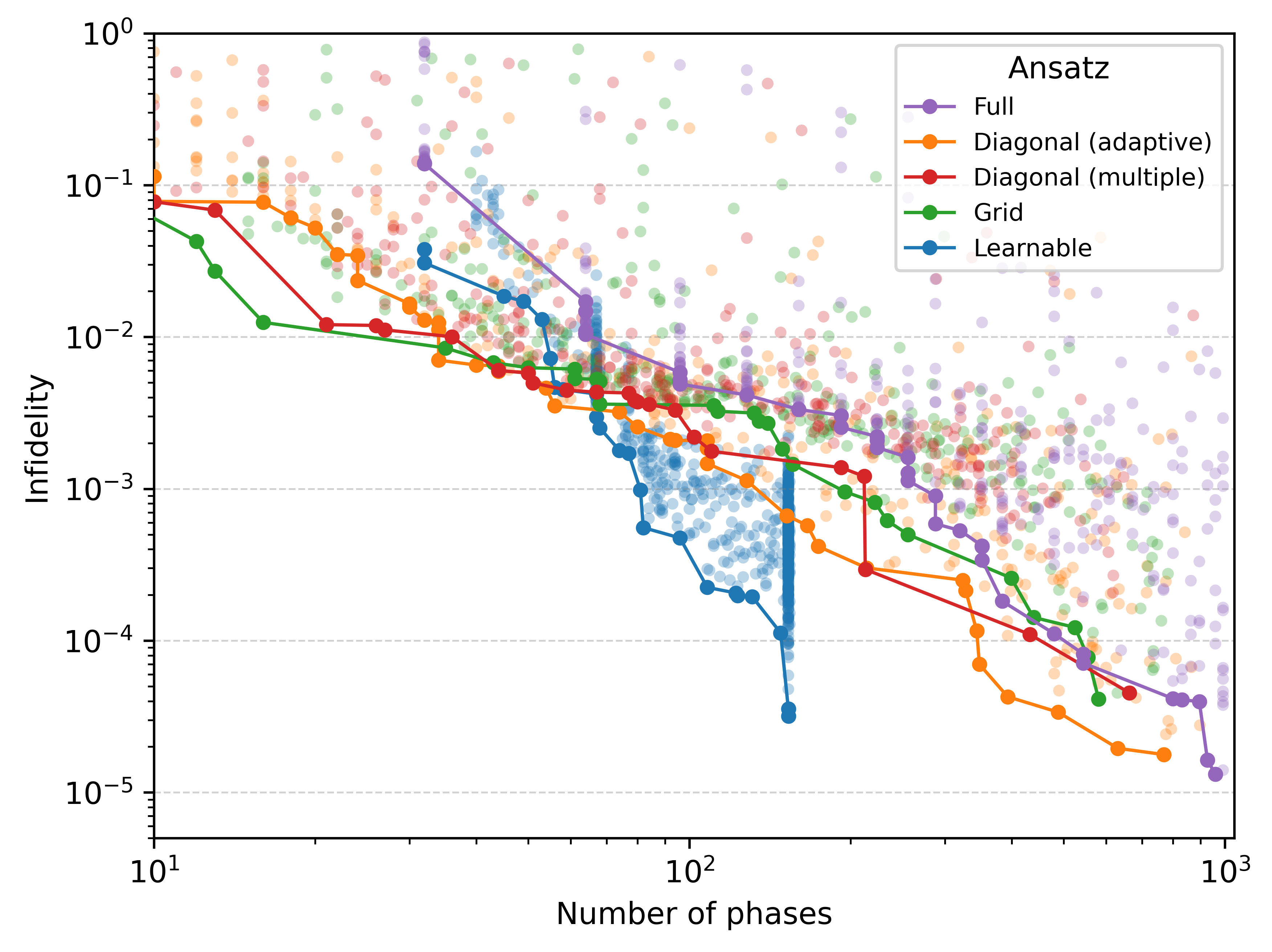}
        \caption{$d=32$.}
        \label{fig:pareto_fourier_d32}
    \end{subfigure}

    \caption{Pareto frontiers comparing the trade-off between infidelity
    and the number of SNAP phases for the Fourier-5 target state.}
    \label{fig:pareto_fourier}
\end{figure}

\begin{figure}[t]
    \centering

    \begin{subfigure}[t]{0.48\linewidth}
        \centering
        \includegraphics[width=\linewidth]
        {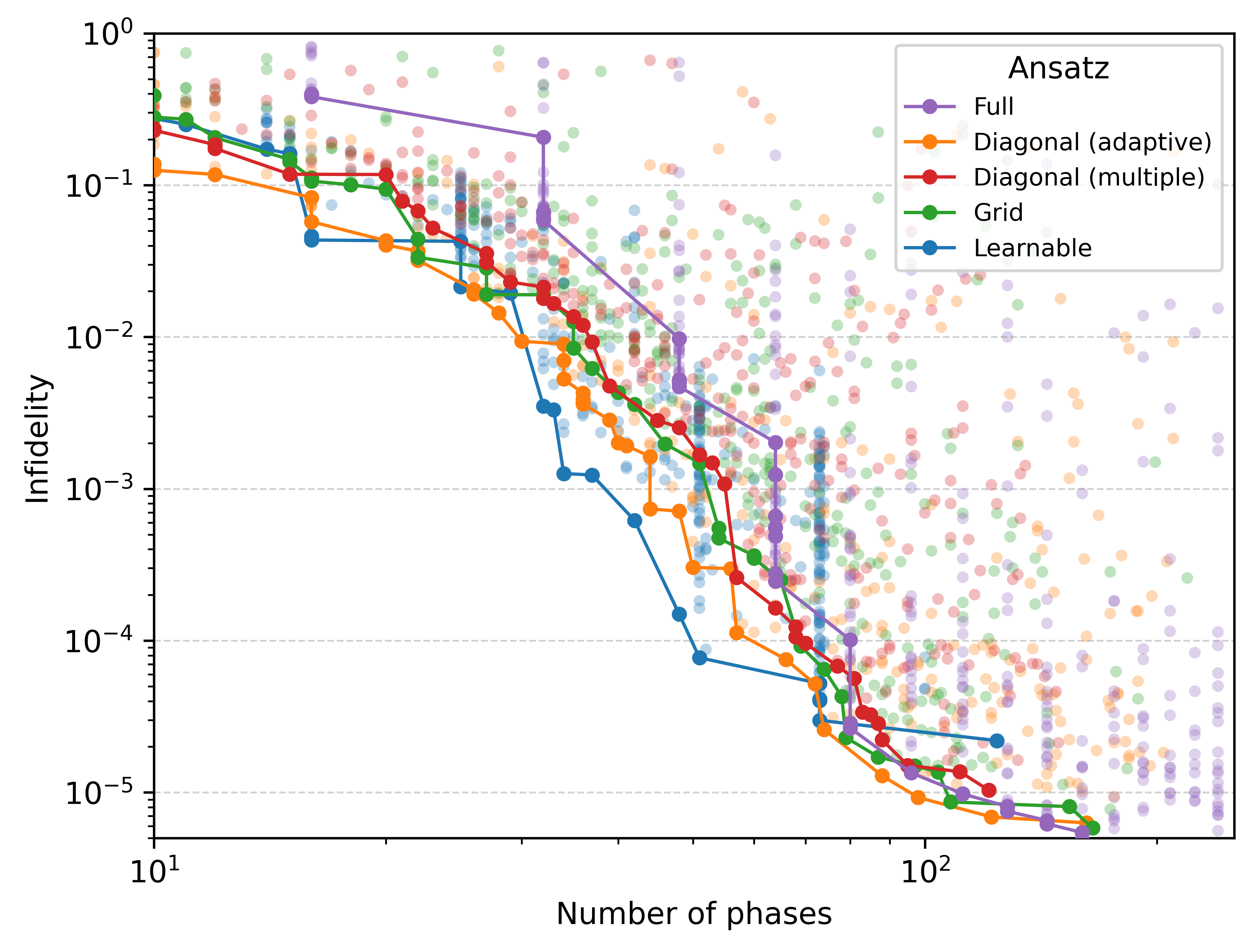}
        \caption{$d=16$.}
        \label{fig:pareto_haar_d16}
    \end{subfigure}
    \hfill
    \begin{subfigure}[t]{0.48\linewidth}
        \centering
        \includegraphics[width=\linewidth]
        {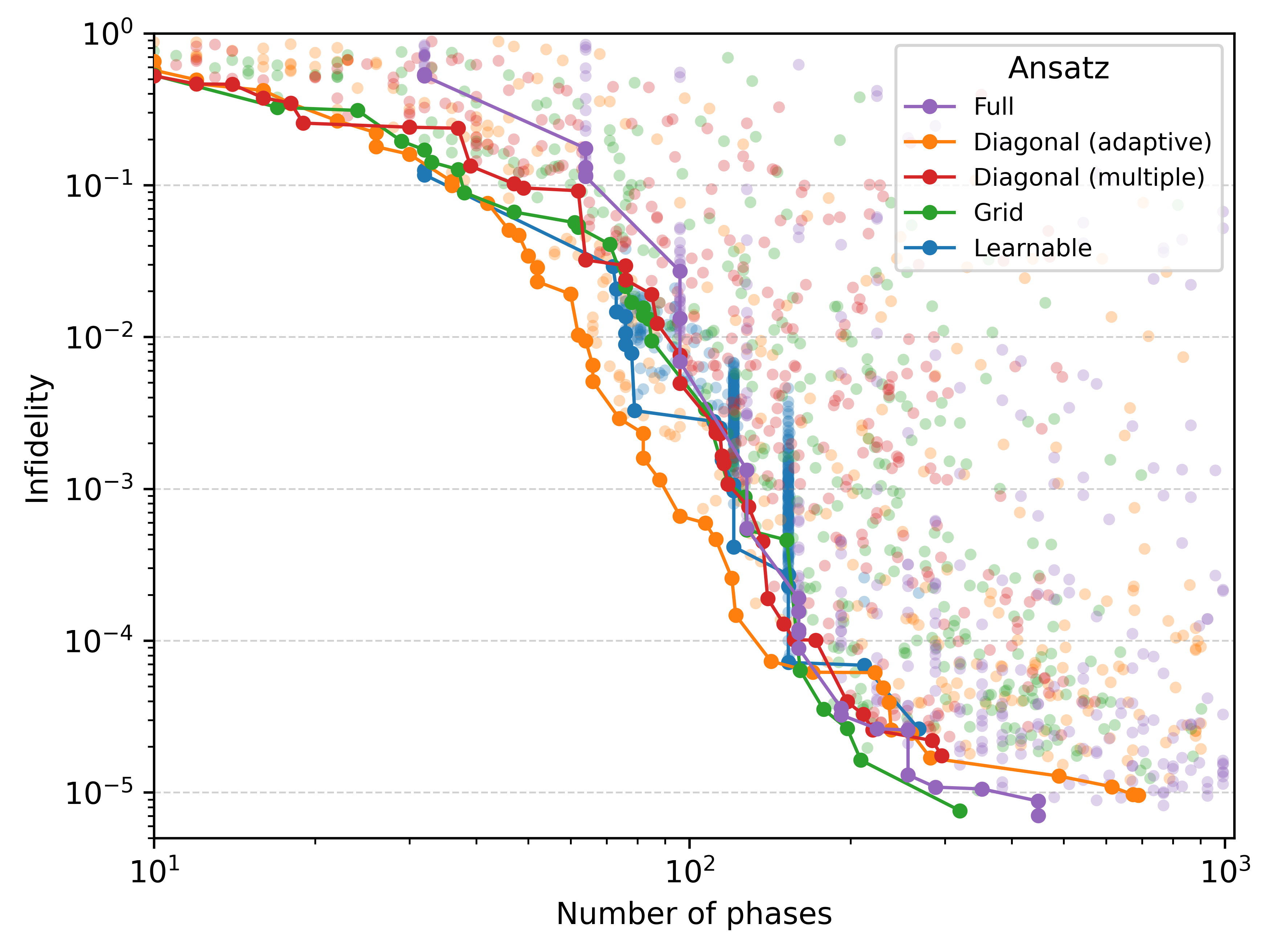}
        \caption{$d=32$.}
        \label{fig:pareto_haar_d32}
    \end{subfigure}

    \caption{Pareto frontiers comparing the trade-off between infidelity
    and the number of SNAP phases for the Haar random target state.}
    \label{fig:pareto_haar}
\end{figure}

Figures~\ref{fig:pareto_fourier} and~\ref{fig:pareto_haar} show the Pareto frontiers for the Fourier-5 and Haar random states at $d=16$ and $d=32$. For Fourier-5, \texttt{Learnable} provides a clear advantage at intermediate infidelities, approximately between $10^{-2}$ and $10^{-4}$, where it reaches similar or lower infidelities using fewer SNAP phases. It should be noted that this corresponds broadly to the target infidelities used for the hyperparameter optimization of \texttt{Learnable}. This behavior is visible at both dimensions and is particularly pronounced at $d=32$. For the Haar random state, the advantage is smaller, \texttt{Learnable} is competitive at $d=16$, while at $d=32$ the frontiers largely overlap and no ansatz performs consistently better over the complete range. In general, the differences are less marked at the lowest and highest infidelities. This indicates that \texttt{Learnable} is most effective in the intermediate infidelity region. Its reduced advantage at very low infidelities may be due to the increased difficulty of jointly optimizing the gate parameters and the learned mask when higher accuracy is required.

\subsection{Hypervolume Analysis}
\label{sec:bootstrap_hv}

While the Pareto frontiers show where the differences between the ansatzes occur, their comparison becomes less immediate when several target states and qudit dimensions are considered. We therefore use hypervolume to summarize the overall trade-off between infidelity and the number of SNAP phases. Within the same analysis and qudit dimension, a larger hypervolume indicates a better trade-off.

Following~\citet{dacrema2026sparse}, we first compute the hypervolume over the complete range of infidelities explored during the experiments. Since the Pareto frontiers show that \texttt{Learnable} is most competitive in the intermediate region, we also perform a second analysis including only configurations satisfying $I\leq10^{-3}$. The two hypervolumes are computed independently, and their absolute values should therefore not be compared across the two analyses. Comparisons are made only between ansatzes within the same qudit dimension and panel. Tables~\ref{tab:hv_fourier} and~\ref{tab:hv_haar} report the mean hypervolume and the corresponding $95\%$ confidence intervals obtained from 1000 bootstrap samples.

Over the complete range, \texttt{Diagonal (adaptive)} obtains the highest mean hypervolume for both target states and all the considered dimensions. For Fourier-5, \texttt{Learnable} remains close to \texttt{Diagonal (adaptive)} at $d=16$ and $d=32$, with overlapping confidence intervals, but does not improve on its mean hypervolume. For the Haar random state, the intervals also overlap at $d=8$ and $d=16$. At $d=32$, however, \texttt{Learnable} obtains a clearly lower value, with $32.1\,[30.2,33.8]$ compared with $36.2\,[35.8,36.8]$ for \texttt{Diagonal (adaptive)}.

The restricted analysis gives a different result for Fourier-5. \texttt{Learnable} obtains the highest mean hypervolume at all three dimensions. At $d=8$, its confidence interval overlaps those of the comparison ansatzes, so the small difference in the mean does not indicate a clear advantage. The improvement is more marked at $d=16$, where \texttt{Learnable} obtains $55.0\,[53.0,57.5]$ compared with $48.9\,[48.2,49.7]$ for \texttt{Diagonal (adaptive)}, and at $d=32$, where it obtains $41.7\,[39.1,43.6]$ compared with $32.8\,[32.1,33.7]$. In both cases, the confidence intervals do not overlap. For the Haar random state, restricting the analysis does not produce the same consistent improvement. At $d=8$, \texttt{Learnable} has a slightly higher mean than \texttt{Diagonal (adaptive)}, $73.1$ compared with $72.9$, but their confidence intervals overlap substantially. At $d=16$, \texttt{Diagonal (adaptive)} has the higher mean, although the intervals again overlap. At $d=32$, \texttt{Diagonal (adaptive)} remains clearly better, obtaining $46.7\,[46.0,47.5]$ compared with $41.6\,[38.8,44.3]$ for \texttt{Learnable}.

Overall, the two analyses show that \texttt{Learnable} is most competitive in the accuracy region where the Pareto frontiers indicate its main advantage. Although \texttt{Diagonal (adaptive)} performs better over the complete infidelity range, restricting the analysis to $I\leq10^{-3}$ reveals clear improvements for Fourier-5, particularly at $d=16$ and $d=32$. The results for the Haar random state are more mixed, indicating that the benefit of learning the mask depends on the target state.

\begin{table*}[t]
    \centering
    \small

    \begin{subtable}[t]{0.95\textwidth}
        \centering
        \resizebox{\linewidth}{!}{
            \begin{tabular}{l|c|c|c|c|c}
\toprule
d & \texttt{Full} & \texttt{Diagonal (adaptive)} & \texttt{Diagonal (multiple)} & \texttt{Grid} & \texttt{Learnable} \\
\midrule
8 & 51.9 $[51.5, 52.3]$ & \textbf{57.5 $[56.9, 58.0]$} & 54.1 $[53.3, 54.9]$ & 57.0 $[56.4, 57.6]$ & 54.6 $[52.6, 56.9]$ \\
16 & 40.2 $[39.8, 40.7]$ & \textbf{46.6 $[46.1, 47.2]$} & 43.3 $[42.6, 44.2]$ & 44.9 $[44.3, 45.7]$ & 44.8 $[43.3, 46.6]$ \\
32 & 27.4 $[27.0, 27.9]$ & \textbf{35.2 $[34.6, 35.9]$} & 32.9 $[32.4, 33.5]$ & 32.3 $[31.8, 32.9]$ & 34.8 $[33.1, 36.3]$ \\
% 64 & 19.6 $[18.7, 20.8]$ & \textbf{29.9 $[28.8, 30.9]$} & 28.1 $[27.5, 28.9]$ & 29.1 $[28.5, 29.8]$ & -- \\
\bottomrule
\end{tabular}

        }
        \caption{Complete explored infidelity range.}
        \label{tab:hv_fourier_all}
    \end{subtable}
    \hfill
    \begin{subtable}[t]{0.95\textwidth}
        \centering
        \resizebox{\linewidth}{!}{
            \begin{tabular}{l|c|c|c|c|c}
\toprule
d & \texttt{Full} & \texttt{Diagonal (adaptive)} & \texttt{Diagonal (multiple)} & \texttt{Grid} & \texttt{Learnable} \\
\midrule
8 & 64.9 $[64.4, 65.6]$ & 68.4 $[67.7, 69.2]$ & 65.2 $[64.2, 66.4]$ & 66.4 $[65.7, 67.2]$ & \textbf{69.9 $[66.7, 72.8]$} \\
16 & 46.6 $[46.0, 47.2]$ & 48.9 $[48.2, 49.7]$ & 41.4 $[40.5, 42.6]$ & 46.2 $[45.4, 47.2]$ & \textbf{55.0 $[53.0, 57.5]$} \\
32 & 26.8 $[26.3, 27.4]$ & 32.8 $[32.1, 33.7]$ & 27.1 $[26.4, 27.9]$ & 27.8 $[27.2, 28.7]$ & \textbf{41.7 $[39.1, 43.6]$} \\
% 64 & 14.5 $[13.4, 16.1]$ & \textbf{28.8 $[27.4, 30.2]$} & 21.7 $[21.0, 22.8]$ & 24.2 $[23.4, 25.1]$ & -- \\
\bottomrule
\end{tabular}

        }
        \caption{Configurations satisfying $I\leq10^{-3}$.}
        \label{tab:hv_fourier_restricted}
    \end{subtable}

    \caption{Bootstrap hypervolume for the Fourier-5 target state, computed
    using 1000 bootstrap samples. The two analyses are computed independently,
    so their absolute hypervolume values are not directly comparable. Values
    should be compared across ansatzes within the same row and panel.}
    \label{tab:hv_fourier}
\end{table*}

\begin{table*}[t]
    \centering
    \small

    \begin{subtable}[t]{0.95\textwidth}
        \centering
        \resizebox{\linewidth}{!}{
            \begin{tabular}{l|c|c|c|c|c}
\toprule
d & \texttt{Full} & \texttt{Diagonal (adaptive)} & \texttt{Diagonal (multiple)} & \texttt{Grid} & \texttt{Learnable} \\
\midrule
8 & 49.1 $[48.4, 50.0]$ & \textbf{53.9 $[53.2, 54.7]$} & 52.3 $[51.2, 53.4]$ & 53.1 $[51.9, 54.1]$ & 53.1 $[49.8, 55.5]$ \\
16 & 41.1 $[40.6, 41.7]$ & \textbf{45.3 $[44.6, 46.0]$} & 42.5 $[41.6, 43.6]$ & 44.1 $[43.5, 44.8]$ & 44.0 $[41.9, 46.7]$ \\
32 & 32.6 $[32.2, 33.0]$ & \textbf{36.2 $[35.8, 36.8]$} & 33.7 $[32.9, 34.6]$ & 35.3 $[34.5, 36.1]$ & 32.1 $[30.2, 33.8]$ \\
% 64 & 22.2 $[21.2, 23.4]$ & \textbf{26.5 $[25.1, 28.0]$} & 24.6 $[23.4, 26.1]$ & 24.8 $[23.5, 26.8]$ & -- \\
\bottomrule
\end{tabular}

        }
        \caption{Complete explored infidelity range.}
        \label{tab:hv_haar_all}
    \end{subtable}
    \hfill
    \begin{subtable}[t]{0.95\textwidth}
        \centering
        \resizebox{\linewidth}{!}{
            \begin{tabular}{l|c|c|c|c|c}
\toprule
d & \texttt{Full} & \texttt{Diagonal (adaptive)} & \texttt{Diagonal (multiple)} & \texttt{Grid} & \texttt{Learnable} \\
% d &  &  &  &  &  \\
\midrule
8 & 67.7 $[66.6, 68.9]$ & 72.9 $[71.9, 74.1]$ & 70.5 $[68.8, 72.2]$ & 70.8 $[69.1, 72.2]$ & \textbf{73.1 $[67.5, 76.2]$} \\
16 & 56.9 $[56.1, 57.7]$ & \textbf{59.7 $[58.7, 60.8]$} & 54.9 $[53.6, 56.6]$ & 57.6 $[56.6, 58.6]$ & 58.3 $[55.2, 62.5]$ \\
32 & 45.0 $[44.5, 45.6]$ & \textbf{46.7 $[46.0, 47.5]$} & 42.5 $[41.5, 43.8]$ & 45.0 $[43.9, 46.2]$ & 41.6 $[38.8, 44.3]$ \\
% 64 & 28.5 $[26.9, 30.5]$ & \textbf{30.6 $[28.4, 33.0]$} & 29.1 $[27.1, 31.6]$ & 26.7 $[25.0, 28.7]$ & -- \\
\bottomrule
\end{tabular}

        }
        \caption{Configurations satisfying $I\leq10^{-3}$.}
        \label{tab:hv_haar_restricted}
    \end{subtable}

    \caption{Bootstrap hypervolume for the Haar random target state, computed
    using 1000 bootstrap samples. The two analyses are computed independently,
    so their absolute hypervolume values are not directly comparable. Values
    should be compared across ansatzes within the same row and panel.}
    \label{tab:hv_haar}
\end{table*}

\section{Conclusions}

This work investigates how learning based methods can be used to select the SNAP phases required for bosonic state preparation. Building on the heuristic sparse ansatzes introduced by~\citet{dacrema2026sparse}, we extended the \texttt{Diagonal (adaptive)} ansatz with a differentiable mask that is optimized together with the gate parameters. The method starts from the phases allowed by the \texttt{Diagonal (adaptive)} structure and learns which of them should be retained.

We evaluated the proposed \texttt{Learnable} ansatz on Fourier-5 and Haar random target states for qudit dimensions up to $d=32$. The comparison considered the minimum number of phases required to reach different target infidelities, the corresponding Pareto frontiers, and the hypervolume over both the complete explored range and the region satisfying $I\leq10^{-3}$.

The results show that \texttt{Learnable} can reduce the number of SNAP phases required to reach a given infidelity. The improvements are clearest for Fourier-5, particularly at $d=16$ and $d=32$, where the method provides a better trade-off in the intermediate infidelity region. This is also reflected in the restricted hypervolume analysis, where \texttt{Learnable} obtains the highest mean hypervolume for Fourier-5 at all considered dimensions. For the Haar random state, the results are more mixed: \texttt{Learnable} remains competitive and improves several individual configurations, but \texttt{Diagonal (adaptive)} performs better in some dimensions and regions of the frontier.

Overall, the results indicate that learning which phases to retain can improve predefined sparse ansatzes, although the benefit depends on the target state and on the range of infidelities considered. The current method can only remove phases from the initial \texttt{Diagonal (adaptive)} structure. Future work may therefore consider learning phase locations over a broader initial structure, different mask architectures and studying its behavior under realistic noise and hardware constraints.

\begin{acknowledgments}

We acknowledge that the scholarship of Marco Pennazio has been funded by ABB S.p.A. 
We acknowledge ISCRA for awarding this project access to the LEONARDO supercomputer, owned by the EuroHPC Joint Undertaking, hosted by CINECA (Italy).
\end{acknowledgments}

\section*{Declaration on Generative AI}
During the preparation of this work, the authors used GPT-5.6 Sol in order to paraphrase and reword, as well as for grammar and spelling checks. After using these tools, the authors reviewed and edited the content as needed and take full responsibility for the publication's content.

\bibliography{sample-ceur}

@article{dacrema2026sparse,
  author       = {Maurizio {Ferrari Dacrema} and
                  Do\u{g}a Murat K\"urk\c{c}\"uo\u{g}lu and
                  Andy Li and
                  Tanay Roy and
                  Silvia Zorzetti},
  title        = {Sparse Phase Ansatzes for Resource-Efficient Qudit State Preparation via the {SNAP}-Displacement Protocol},
  year         = {2026},
  eprinttype    = {arXiv},
  eprint       = {2603.12203},
}

@article{chakram2022multimode,
  title={Multimode photon blockade},
  author={Chakram, Srivatsan and He, Kevin and Dixit, Akash V and Oriani, Andrew E and Naik, Ravi K and Leung, Nelson and Kwon, Hyeokshin and Ma, Wen-Long and Jiang, Liang and Schuster, David I},
  journal={Nature Physics},
  volume={18},
  number={8},
  pages={879--884},
  year={2022},
  publisher={Nature Publishing Group UK London},
  doi={10.1038/s41567-022-01630-y}
}

@article{krayzman2024superconducting,
  title={Superconducting quantum memory with a suspended coaxial resonator},
  author={Krayzman, Lev and Lei, Chan U and Ganjam, Suhas and Teoh, James and Frunzio, Luigi and Schoelkopf, Robert J},
  journal={Applied Physics Letters},
  volume={124},
  number={20},
  year={2024},
  publisher={AIP Publishing},
  doi={10.1063/5.0203906}
}

@article{PRXQuantum.3.010335,
  title = {Encoding Qubits in Multimode Grid States},
  author = {Royer, Baptiste and Singh, Shraddha and Girvin, S.M.},
  journal = {PRX Quantum},
  volume = {3},
  issue = {1},
  pages = {010335},
  numpages = {40},
  year = {2022},
  month = {Mar},
  publisher = {American Physical Society},
  doi = {10.1103/PRXQuantum.3.010335}
}

@article{PRXQuantum.4.020342,
  title = {Qubit-Oscillator Concatenated Codes: Decoding Formalism and Code Comparison},
  author = {Xu, Yijia and Wang, Yixu and Kuo, En-Jui and Albert, Victor V.},
  journal = {PRX Quantum},
  volume = {4},
  issue = {2},
  pages = {020342},
  numpages = {22},
  year = {2023},
  month = {Jun},
  publisher = {American Physical Society},
  doi = {10.1103/PRXQuantum.4.020342}
}

@article{fosel2020efficient,
  title={Efficient cavity control with SNAP gates},
  author={F{\"o}sel, Thomas and Krastanov, Stefan and Marquardt, Florian and Jiang, Liang},
  journal={arXiv preprint arXiv:2004.14256},
  year={2020}
}

@article{SNAP2015PRL,
  title = {Cavity State Manipulation Using Photon-Number Selective Phase Gates},
  author = {Heeres, Reinier W. and Vlastakis, Brian and Holland, Eric and Krastanov, Stefan and Albert, Victor V. and Frunzio, Luigi and Jiang, Liang and Schoelkopf, Robert J.},
  journal = {Phys. Rev. Lett.},
  volume = {115},
  issue = {13},
  pages = {137002},
  numpages = {5},
  year = {2015},
  month = {Sep},
  publisher = {American Physical Society},
  doi = {10.1103/PhysRevLett.115.137002},
  url = {https://link.aps.org/doi/10.1103/PhysRevLett.115.137002}
}

@article{PhysRevA.92.040303,
  title = {Universal control of an oscillator with dispersive coupling to a qubit},
  author = {Krastanov, Stefan and Albert, Victor V. and Shen, Chao and Zou, Chang-Ling and Heeres, Reinier W. and Vlastakis, Brian and Schoelkopf, Robert J. and Jiang, Liang},
  journal = {Phys. Rev. A},
  volume = {92},
  issue = {4},
  pages = {040303},
  numpages = {5},
  year = {2015},
  month = {Oct},
  publisher = {American Physical Society},
  doi = {10.1103/PhysRevA.92.040303},
  url = {https://link.aps.org/doi/10.1103/PhysRevA.92.040303}
}

@article{PRXQuantum.3.030301,
  title = {Robust Preparation of Wigner-Negative States with Optimized SNAP-Displacement Sequences},
  author = {Kudra, Marina and Kervinen, Mikael and Strandberg, Ingrid and Ahmed, Shahnawaz and Scigliuzzo, Marco and Osman, Amr and Lozano, Daniel P\'erez and Thol\'en, Mats O. and Borgani, Riccardo and Haviland, David B. and Ferrini, Giulia and Bylander, Jonas and Kockum, Anton Frisk and Quijandr\'{\i}a, Fernando and Delsing, Per and Gasparinetti, Simone},
  journal = {PRX Quantum},
  volume = {3},
  issue = {3},
  pages = {030301},
  numpages = {12},
  year = {2022},
  month = {Jul},
  publisher = {American Physical Society},
  doi = {10.1103/PRXQuantum.3.030301},
  url = {https://link.aps.org/doi/10.1103/PRXQuantum.3.030301}
}

@article{RevModPhys.97.021003,
  title = {Colloquium: Qudits for decomposing multiqubit gates and realizing quantum algorithms},
  author = {Kiktenko, Evgeniy O. and Nikolaeva, Anastasiia S. and Fedorov, Aleksey K.},
  journal = {Rev. Mod. Phys.},
  volume = {97},
  issue = {2},
  pages = {021003},
  numpages = {26},
  year = {2025},
  month = {Jun},
  publisher = {American Physical Society},
  doi = {10.1103/RevModPhys.97.021003},
  url = {https://link.aps.org/doi/10.1103/RevModPhys.97.021003}
}

@article{bornman2025benchmarking,
  title={Benchmarking the performance of a high-Q cavity qudit using random unitaries},
  author={Bornman, Nicholas and Roy, Tanay and Job, Joshua A and Anand, Namit and Perdue, Gabriel N and Zorzetti, Silvia and Alam, M Sohaib},
  journal={Quantum Science and Technology},
  volume={10},
  number={2},
  pages={025062},
  year={2025},
  publisher={IOP Publishing}
}

@article{Heeres2017,
  author  = {Reinier W. Heeres and Philip Reinhold and Nissim Ofek and Luigi Frunzio and Liang Jiang and Michel H. Devoret and Robert J. Schoelkopf},
  title   = {Implementing a universal gate set on a logical qubit encoded in an oscillator},
  journal = {Nature Communications},
  year    = {2017},
  volume  = {8},
  number  = {1},
  pages   = {94},
  doi     = {10.1038/s41467-017-00045-1},
  url     = {https://doi.org/10.1038/s41467-017-00045-1},
  issn    = {2041-1723}
}

@article{job2023efficient,
  title={Efficient, direct compilation of SU (N) operations into SNAP \& Displacement gates},
  author={Job, Joshua},
  journal={arXiv preprint arXiv:2307.11900},
  year={2023}
}

@conference{Roy2024Qudit,
  author       = {Roy, Tanay and Kim, Taeyoon and Romanenko, Alexander and Grassellino, Anna},
  title        = {Qudit-based quantum computing with SRF cavities at Fermilab},
  doi          = {10.22323/1.453.0127},
  url          = {https://www.osti.gov/biblio/2318671},
  journal      = {PoS},
  volume       = {LATTICE2023},
  place        = {United States},
  organization = {Fermi National Accelerator Laboratory (FNAL), Batavia, IL (United States)},
  year         = {2024},
  month        = {01}}

@ARTICLE{10.3389/fphy.2020.589504,
AUTHOR={Wang, Yuchen  and Hu, Zixuan  and Sanders, Barry C.  and Kais, Sabre },        
TITLE={Qudits and High-Dimensional Quantum Computing},       
JOURNAL={Frontiers in Physics},      
VOLUME={Volume 8 - 2020},
YEAR={2020},
URL={https://www.frontiersin.org/journals/physics/articles/10.3389/fphy.2020.589504},
DOI={10.3389/fphy.2020.589504},
ISSN={2296-424X},
}

@article{hypervolume1,
author = {Guerreiro, Andreia P. and Fonseca, Carlos M. and Paquete, Lu\'{\i}s},
title = {The Hypervolume Indicator: Computational Problems and Algorithms},
year = {2021},
issue_date = {July 2022},
publisher = {Association for Computing Machinery},
address = {New York, NY, USA},
volume = {54},
number = {6},
issn = {0360-0300},
url = {https://doi.org/10.1145/3453474},
doi = {10.1145/3453474},
journal = {ACM Comput. Surv.},
month = jul,
articleno = {119},
numpages = {42},
}

@article{hypervolume2,
title = {Hypervolume-based multiobjective optimization: Theoretical foundations and practical implications},
journal = {Theoretical Computer Science},
volume = {425},
pages = {75-103},
year = {2012},
note = {Theoretical Foundations of Evolutionary Computation},
issn = {0304-3975},
doi = {https://doi.org/10.1016/j.tcs.2011.03.012},
url = {https://www.sciencedirect.com/science/article/pii/S0304397511002428},
author = {Anne Auger and Johannes Bader and Dimo Brockhoff and Eckart Zitzler},
}

@article{Bazgan2022,
  author  = {Bazgan, Cristina and Ruzika, Stefan and Thielen, Clemens and Vanderpooten, Daniel},
  title   = {The Power of the Weighted Sum Scalarization for Approximating Multiobjective Optimization Problems},
  journal = {Theory of Computing Systems},
  year    = {2022},
  volume  = {66},
  number  = {1},
  pages   = {395--415},
  doi     = {10.1007/s00224-021-10066-5},
  url     = {https://doi.org/10.1007/s00224-021-10066-5},
  issn    = {1433-0490}
}
\end{document}